\documentclass[aps,prl,showpacs,twocolumn,floatfix,unsortedaddress]{revtex4}
\usepackage{amssymb}

\usepackage{amsfonts}
\usepackage{graphicx}

\begin{document}
\title{Synchronized chaos in networks of simple units}
\author{Frank Bauer}
\email{bauer@mis.mpg.de}
\author{Fatihcan M. Atay}
\email{fatay@mis.mpg.de}
\affiliation{Max Planck Institute for Mathematics in the Sciences,
Inselstrasse 22, D-04103 Leipzig, Germany}
\author{J\"urgen Jost}
\email{jjost@mis.mpg.de} \affiliation{Max Planck Institute for
Mathematics in the Sciences, Inselstrasse 22, D-04103 Leipzig,
Germany}
\address{ Santa Fe Institute for the Sciences of
Complexity, Santa Fe, NM 87501, USA}

\noindent
Preprint. Final version in: \\ EPL, \textbf{89} (2010) 20002 \\
 doi: 10.1029/0295-5075/89/20002
\begin{abstract}We study synchronization of non-diffusively coupled map
networks with arbitrary network topologies, where the connections
between different units are, in general, not symmetric and can carry
both positive and negative weights. We show that, in contrast to
diffusively coupled networks, the synchronous behavior of a
non-diffusively coupled network can be dramatically different from
the behavior of its constituent units. In particular, we show that
chaos can emerge as synchronized behavior although the dynamics of
individual units are very simple. Conversely, individually chaotic
units can display simple behavior when the network synchronizes. We
give a synchronization criterion that depends on the spectrum of the
generalized graph Laplacian, as well as the dynamical properties of
the individual units and the interaction function. This general
result will be applied to coupled systems of  tent and logistic maps
and to two models of neuronal dynamics. Our approach yields an
analytical understanding of how simple model neurons can produce
complex collective behavior through the coordination of their
actions.

\end{abstract}
\maketitle
\section{Introduction}
Dynamical processes in networks, such as synchronization, have been
attracting much interest \cite{Pikovsky-book,Boccaletti02,
Arenas08,Chavez07}. A striking characteristic of many networks is
that they are often formed from very simple units (\textit{e.g.} a
neuron either spikes or is silent, at a certain level of
description) but can collectively exhibit a wide range of dynamics.
A central question is then how dynamically simple units can produce
rich collective dynamical behavior when they are coupled together in
a network. In this letter we offer a solution to this question in
the context of synchronization of coupled map networks.

Previous work on synchronization of coupled maps focused on
diffusive coupling with non-negative weights. However, in
diffusively-coupled networks the synchronized network shows the same
dynamical behavior as one single isolated unit;
thus, no new collective behavior is emerging here. New collective
behavior could, for instance, be produced by time delays, which may
remarkably make it easier for networks to synchronize \cite{Atay04b,
Atay06}.

Here however, instead of  time delays, we consider non-diffusive
coupling schemes. One particular non-diffusive coupling scheme, the
so-called \textit{direct coupling scheme} is motivated by biological
findings (see \cite{Dayan01} and the references therein) and has
been used in studies of amplitude response of coupled oscillators
\cite{Aronson90}, although not investigated as extensively as
diffusive coupling in synchronization research. In this letter, we
use a direct coupling scheme to study the emergence of new
collective dynamical behavior. In particular, we show the emergence
of synchronized chaotic behavior in a network of non-chaotic units.
To our knowledge this is the first time that such a phenomenon is
observed and analyzed in depth in mathematical network models. In
contrast, synchronized chaotic behavior in a network of chaotic
units \cite{Pecora90,Kaneko90} and non-synchronized chaotic behavior
in a network of non-chaotic units \cite{Li04} are well established
phenomena.

A further feature of this work is that we take the succeeding,
typically in the literature neglected, facts into
account. 
Many biological networks share the following two properties
\cite{Dayan01}: $(i)$ The connection structure is, in general, not
symmetric. $(ii)$ The influence of neighboring units can be
excitatory or inhibitory, which is modelled by positive and negative
weights. It is thus essential to incorporate these characteristics
in network models in order to understand the dynamical behavior of
biological networks. Consequently, we consider networks with
arbitrary network topologies, namely, not necessarily symmetrically
coupled networks with possibly both positive and negative weights.
On the other hand, we restrict ourselves to networks of identical
units. We mention, e.g., \cite{Sun09}, as a recent study of
diffusively-coupled units with small parametric variations.

In order to emphasize a general aspect, we consider in the next
section networks with pairwise coupling and present a general
synchronization criterion. Later on, we will focus on directly
coupled networks and study the emergence of new behavior.

\section{Pairwise coupling} In our coupled map network model, each
node is a dynamical system whose evolution is described in discrete
time $t$ by iterations of a scalar map $f$, \textit{i.e.}~by an
equation of the form
\begin{equation} x(t+1)=f(x(t)). \label{map}
\end{equation}
The interconnections are specified by a  weighted, directed graph
$\Gamma$ on $n$ vertices. The weight $w_{ij}$ of the connection from
vertex $j$ to vertex $i$ can be positive, negative or zero. We
assume that the network has no self-loops, that is, $w_{ii}=0$ for
all  $i$. The in-degree of vertex $i$ is
 $d_i = \sum_{j=1}^n w_{ij}$ \footnote{In principle
   there may exist vertices with zero in-degree
    because of cancellations between positive and
negative weights. For simplicity we exclude such vertices here,
since they may prevent the existence of a synchronized solution for
non-diffusively coupled units. The general case is treated in
\cite{Bauer08b}.}. The activity at vertex or unit $i$ at time $t+1$
is given by:
\begin{eqnarray}\label{pairwise}
x_i(t+1) &=& f(x_i(t)) + \frac{\epsilon}{d_i}\sum_{j=1}^n w
_{ij}g(x_i(t),x_j(t)),\\&&\nonumber \qquad\qquad \qquad\qquad
\qquad\, i = 1,...,n ,
\end{eqnarray}
where $f: \mathbb{R} \rightarrow \mathbb{R}$ and $g: \mathbb{R}^2
\rightarrow \mathbb{R}$ are 
differentiable functions with bounded derivatives, and $\epsilon \in
\mathbb{R}$ is the overall coupling strength. The function $f$
describes the dynamical behavior of the individual units whereas $g$
characterizes the interactions between different pairs of units.

\section{Synchronization} We are interested in synchronized
solutions of eq.~(\ref{pairwise}), where the activity of all units
is identical, that is, $x_i(t) = s(t)$ for all $i$ and $t$. It
follows from eq.~(\ref{pairwise}) that a synchronized solution
$s(t)$ satisfies \begin{equation} \label{9} s(t+1) = f(s(t)) +
\epsilon g(s(t),s(t)).
\end{equation}
This equation already shows that the synchronized solution $s(t)$
can be quite different from the dynamical behavior of an isolated
unit described by $f$. By contrast, in diffusive-type coupling,
\textit{i.e.} $g(x,x) = 0$ for all $x$, the interaction $g$ vanishes
when the network is synchronized; therefore, the synchronized
solution is identical to the behavior of the individual units, and
no new dynamics can emerge
from synchronization.

Before we explore different examples of new collective behavior, we
investigate the robustness of the synchronized state against
perturbations.
The network is said to (locally) synchronize if $\lim_{t \rightarrow
\infty} |x_i(t) - x_j(t)| = 0$ for all $i,j$ starting from initial
conditions in some appropriate open set\footnote{For chaotic
synchronization there exist  subtleties concerning this open set and
the exact notion of attraction. These issues are carefully studied
in \cite{Lu07}, but will not be important for the purposes of this
letter.}. The propensity of the network to synchronize depends on
the properties of the functions $f$ and $g$ and the underlying
network structure. The latter can be encoded in terms of the
eigenvalues of the graph Laplacian  $\mathcal{L}$ for directed
weighted graphs, defined as \cite{Bauer08}
\begin{equation}(\mathcal{L})_{ij} := \left\{
\begin{array}{r cl} 1 & & \,\mbox{ if}  \;i = j \mbox{ and $d_i \neq 0$}. \\
 -\frac{w_{ij}}{d_i} & &\begin{array}{l}\mbox{if there is a directed edge }\\ \mbox{from $j$ to $i$ and  $d_i\neq 0$.} \end{array} \\
0& &  \,\mbox{ otherwise.}
\end{array} \right.\end{equation}
We label the eigenvalues of $\mathbf{\mathcal{L}}$ as $
\lambda_1,..., \lambda_n$. Since we assume that the in-degrees are
non-zero,
we may write $\mathcal{L} = I - D^{-1}W,$ where $I$ is the ($n
\times n$) identity matrix,
 $D = \mathrm{diag}\{d_1,..., d_n\}$ is the
diagonal matrix of vertex in-degrees and $W =
\left(w_{ij}\right)_{i,j = 1}^n$ is the weighted adjacency matrix of
the underlying graph. 
Zero is always an eigenvalue of $\mathcal{L}$; we denote it
$\lambda_1=0$, and $u_1=(1,\ldots,1)^\top$ is the corresponding
eigenvector.

Since all components of $u_1$ are identical, perturbations along the
$u_1$-direction again yield a synchronous solution.
To study the remaining directions, we define the \textit{k-th mixed
transverse exponent} $\chi_k$ for $2\leq k \leq n$ as:
\begin{equation} \label{chi_k}\chi_k
:=\overline{\lim}_{T\rightarrow\infty}
\frac{1}{T}\sum_{s=\bar{t}}^{\bar{t}+T-1}\log|h_k(s(t))|,\label{30}
\end{equation} where \begin{eqnarray*}h_k(s(t))&=&f'(s(t)) + \epsilon
\partial_1g(s(t),s(t)) \\&&+ \epsilon
\partial_2g(s(t),s(t))(1-\lambda_k)
\end{eqnarray*}
and $\partial_ig$ denotes the ith partial derivative of $g$ and
$\bar{t}$ is chosen such that $h(s(t))\neq 0$ for all $t>\bar{t}$.
If no such $\bar{t}$ exists we set $\chi_k = - \infty$. Note that
these exponents are evaluated along the synchronous solution
(\ref{9}). They combine the dynamical behavior of the individual
units and the interaction function with the network topology. The
maximal mixed transverse exponent governs the synchronizability of
the network, that is, system (\ref{pairwise}) locally synchronizes
if
\begin{equation} \label{5}
\chi := \max_{k \geq 2} \chi_k < 0.
\end{equation}
This result is rigorously derived in our
companion paper \cite{Bauer08b}. 

\section{Diffusive and direct coupling}
In the sequel, we restrict ourselves to functions $g:\mathbb{R}
\rightarrow \mathbb{R}$. When $g(x_i,x_j) = g(x_j)$, pairwise
coupling reduces to direct coupling, i.e.,
\begin{equation}\label{1}
x_i(t+1) = f(x_i(t)) + \frac{\epsilon}{d_i}\sum_{j=1}^n
w_{ij}g(x_j(t)),\;\; i = 1,...,n
\end{equation}
and the mixed transverse exponent (\ref{chi_k}) reduces to
\begin{equation}\label{Chidirect}  \chi^{direct}_k:=
\lim_{T\rightarrow\infty}
\frac{1}{T}\sum_{s=\bar{t}}^{\bar{t}+T-1}\log|f'(s(t)) + \epsilon
g'(s(t))(1-\lambda_k)|.\end{equation} By rearranging terms on the
right hand side in (\ref{1}) as  \begin{equation} f(x_i(t))
+\epsilon g(x_i(t))+ \frac{\epsilon}{d_i}\sum_{j=1}^n
w_{ij}(g(x_j(t))-g(x_i(t))),
\end{equation}
this becomes formally equivalent to a system of the form
\begin{equation} \label{diffusive}
x_i(t+1) = \phi(x_i(t)) + \frac{\epsilon}{d_i}\sum_{j=1}^n
w_{ij}\gamma(x_j(t),x_i(t)),
\end{equation}
with $\gamma(x,x)=0$ for all $x$, i.e., a diffusively coupled map
network. Thus, the conditions for synchronization of directly
coupled networks (\ref{Chidirect}) can be deduced from the diffusive
coupling case (\ref{diffusive}) of \cite{Pecora98}.
However, the formal equivalence obscures the roles of the system
parameters and the particular coupling functions, which are
important in applications. For instance, in neuronal networks, gap
junctions at electrical synapses provide connections of diffusive
type, whereas chemical synapses provide connections with direct
coupling. The distinction is crucial for understanding the effects
of different types of synapses. As already mentioned,
diffusively-coupled networks have been widely studied. For the
remainder of this work, we restrict ourselves to direct coupling.

The definition of $\chi^{direct}_k$ intertwines the effects of the
resulting synchronized dynamics and the network topology. However,
if $g$ is a multiple of $f$, \textit{i.e.} $g = c f$ for some
constant $c$,
then these effects can be separated as the synchronization condition
(\ref{5}) takes the form
\begin{equation} \label{8} \max_{k\ge 2}\log \left|1
-\frac{\epsilon c}{1+ \epsilon c} \lambda_k \right| + \mu_{(1+
\epsilon c)f}  < 0
\end{equation}
where  \begin{equation}\mu_{(1 + \epsilon c)f} :=
\overline{\lim}_{T\rightarrow\infty}\frac{1}{T}\sum_{s=\bar{t}}^{\bar{t}+T-1}\log|(1+
\epsilon c)f'(s(t))|\end{equation}is the Lyapunov exponent of
$(1+\epsilon c)f(x)$.  Here $\bar{t}$ is chosen such that
$f'(s(t))\neq 0$ for all $t>\bar{t}$. In the sequel, let
$\mathcal{D}(c,r)$ denote the disk in the complex plane centered at
$c$ having radius $r$. It is easy to see that (\ref{8}) is
equivalent to the condition that all eigenvalues, except
$\lambda_1$, are contained in $\mathcal{D}(c^*,r^*)$, where
\begin{equation}\label{c*}c^*
= \frac{1+\epsilon c}{\epsilon c}\end{equation}and
\begin{equation}\label{r*} r^* = \left|c^*\right| \exp(-\mu_{(1+
\epsilon c)f}).\end{equation} If, for example, the synchronized
solution $(1 + \epsilon c)f(x)$ is chaotic (\textit{i.e.}~has a
positive Lyapunov exponent), then the first term in (\ref{8})
has to be sufficiently negative to compensate the positive Lyapunov
exponent in order to ensure that the system (\ref{1}) locally
synchronizes. This in turn requires that the eigenvalues $\lambda_k$
for $k\ge 2$
 be bounded away from zero, and the coupling strength
$\epsilon$  lie in an appropriate interval.


\section{Coupled tent maps} Before turning to biologically
motivated functions $f$ and $g$, we demonstrate the emergence of
synchronized chaotic behavior for the case of the tent map where an
analytical treatment is possible. The tent map is given by
\begin{equation}T_\rho(x)= \left\{
\begin{array}{l c l} \rho x, & \mbox{if} & x < \frac{1}{2} \\
 \rho(1-x), &\mbox{if}& x \geq \frac{1}{2},
\end{array} \right. \end{equation}
for  $\rho \in [0,2]$. Its Lyapunov exponent is $\log\rho$; thus, it
is chaotic for $\rho
>1$.  Let
$f(x) = T_a(x)$ and $g(x) = T_b(x)$ with $ 0 < a,b <2$ and choose
the coupling constant $ \epsilon = (\tau - a)/b$. By choosing
different values for the target value $\tau$ we can generate
different synchronized dynamical behavior $s(t+1) = T_\tau(s(t))$
whose Lyapunov exponent equals $\log\tau$. Since the absolute value
of the derivative of $T_\tau(x)$ is constant, from (\ref{8}) we have
that the system (\ref{1}) locally synchronizes if
\begin{equation} \label{6}\left|\tau-(\tau-a)\lambda_k \right| < 1,
\end{equation}  for $k = 2,..., n$. For example, for $\tau = 2$ and $a = b =
1/2$, the synchronized dynamics is chaotic although the individual
units are not. Furthermore, in this case condition (\ref{6}) is
satisfied  if all eigenvalues of the graph Laplacian, except
$\lambda_1$, are contained in $\mathcal{D}(4/3,2/3)$.


\section{Synchronized chaos in neuronal networks} We now apply the
foregoing ideas to models of neuronal networks.

We point out that the equations usually considered in neural network
theory,
\begin{equation} \label{y} y_i(t+1) = f\left(y_i(t) + \frac{\epsilon}{d_i}\sum_{j=1}^n
w_{ij}\varphi(y_j(t))\right) \; ;\; i = 1,...,n,
\end{equation}
can be derived from (\ref{1}) with $\varphi = g \circ f^{-1}$ and
$y_i(t) = f(x_i(t))$. Thus the dynamics of $y_i(t)$ are determined
by the dynamics of $x_i(t)$ and hence our results also apply to
networks given in the form (\ref{y}).

A neuronal network consists of neurons linked by synaptic
connections, which are directed and weighted. For an excitatory
synapse the weight is positive, and the presynaptic neuron increases
the activity  of the postsynaptic neuron according to its weight,
whereas for an inhibitory one the weight is negative, and the
postsynaptic activity is decreased.
\subsection{Leaky neuron model}
\begin{figure}
\includegraphics[width = 7cm]{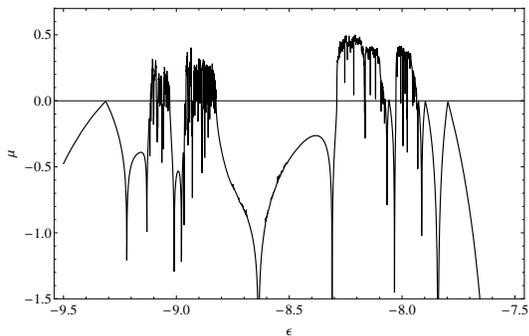} \caption{Lyapunov exponent of
eq.~(\ref{4}) for the parameter values $\gamma = 0.3$, $\kappa =
20$,  $\Theta = 4$.}\label{Fig.2}
\end{figure}

\begin{figure}
\begin{center}
\includegraphics[width = 7cm]{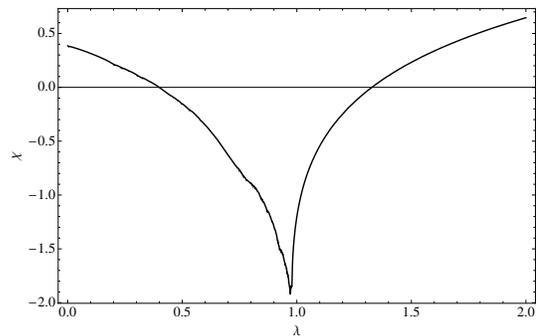} \caption{\label{Fig.6}Mixed
transverse exponent $\chi$ as a function of $\lambda$ for the
dynamics (\ref{4}) and the parameter values  $\gamma = 0.3$, $\kappa
= 20$, $\Theta = 4$, $\epsilon = - 8$. }
\end{center}
\end{figure}

\begin{figure}
\begin{center}
{\includegraphics[width = 7cm]{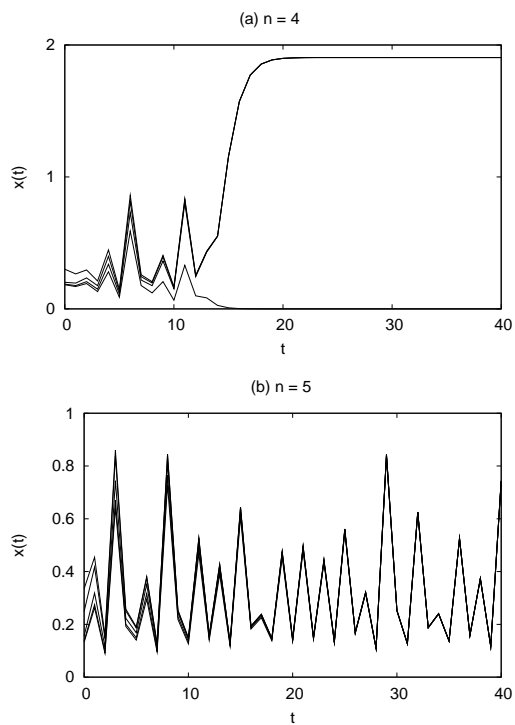}}
\end{center}
\caption{\label{Fig.7}Dynamics of all-to-all coupled leaky neurons
starting from random initial conditions. The 4-neuron system (a)
does not synchronize, but the dynamics remain simple, whereas the
5-neuron system (b) synchronizes, and the synchronous solution is
chaotic. The parameter values are as in fig.~\ref{Fig.6}. }
\end{figure}

In this model, the individual dynamics is governed by (\ref{map})
with $f(x) = \gamma x + \Theta$, where $\gamma\in(0,1)$ represents
dissipation and $\Theta$ is a bias term, which could also include
\textit{e.g.}~an external input.
The interactions between the neurons are modeled by the sigmoidal
function $g(x) = \sigma_\kappa(x)$, where
$$\sigma_\kappa(x) = \frac{1}{1+ \exp(-\kappa x)} -
\frac{1}{2},$$ with $\kappa > 0$.
 The resulting synchronized
solution satisfies
\begin{equation} \label{4} s(t+1) = \gamma s(t) + \Theta  +
\epsilon \sigma_\kappa(s(t)).
\end{equation}
This is a generalization of the dynamics considered for $\kappa = 1$
in \cite{Pasemann97}. In fig.~\ref{Fig.2} the Lyapunov exponent of
eq.~(\ref{4}) is plotted for a set of parameter values. Although the
dynamical behavior of the individual units is very simple (there is
a globally attracting fixed point), the collective behavior can be
non-trivial and even chaotic. Note that dynamical behavior can be
controlled by varying the coupling coefficient $\epsilon$. We now
fix $\epsilon=-8$ so that the synchronized behavior is chaotic. In
fig.~\ref{Fig.6} the mixed transverse exponent $\chi$ is plotted as
a function of $\lambda$, which is here taken to be real for
simplicity of graphical depiction.
The figure shows that the network locally  synchronizes if
approximately
\begin{equation}
\label{7} 0.4 \leq \lambda_k \leq  1.3 \quad \mbox{for } k \ge 2,
\end{equation}
since in that case $\chi_k <0$ for all $k\ge2$. We illustrate the
dynamics in an all-to-all coupled network of leaky neurons. The
eigenvalues fall into the range given by (\ref{7}) when $n>4$. To
see this, recall \cite{Chung97} that the Laplacian of an all-to-all
coupled network on $n$ units has one eigenvalue equal to zero and
all other eigenvalues equal to
\begin{equation} \label{all-to-all}
\lambda_k = n/(n-1), \quad k\ge 2.
\end{equation}
Hence, globally coupled networks having more than four vertices
should synchronize to a common trajectory, which, according to
fig.~\ref{Fig.2}, is chaotic, whereas smaller networks do not
synchronize. This is confirmed by the simulation results of
fig.~\ref{Fig.7}.


\subsection{Sigmoidal neuron model}

\begin{figure}
\begin{center}
\includegraphics[width = 8cm]{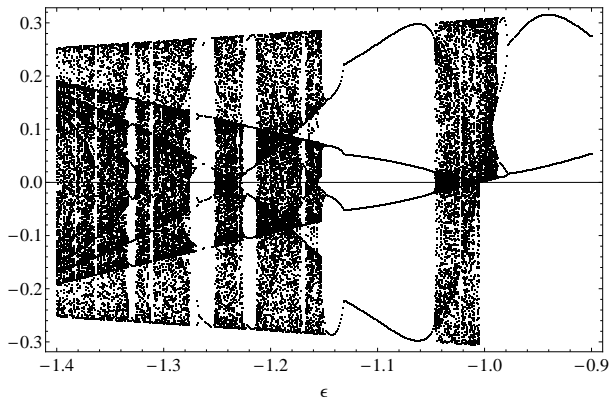} \caption{\label{Fig.3}
Bifurcation diagram of (\ref{3}) for the parameter values $\alpha =
100, \beta = 20$.}
\end{center}
\end{figure}

\begin{figure}
\begin{center}
\includegraphics[width = 6cm]{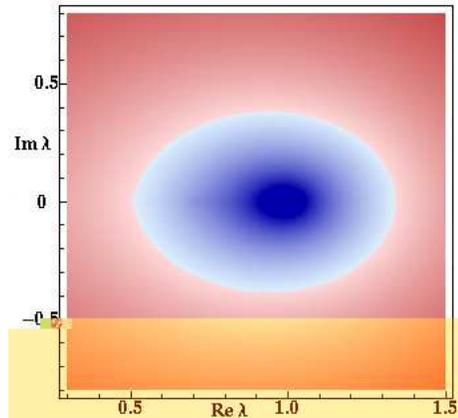} \caption{\label{Fig.5}
The mixed transverse exponent $\chi$ as a function of $\lambda$ for
the dynamics (\ref{3}) and the parameter values $\alpha = 100$,
$\beta = 20$, $\epsilon = - 1$. Blue indicates negatives values of
$\chi$ and red positives values, respectively.}
\end{center}
\end{figure}

\begin{figure}
\begin{center}
\includegraphics[width = 8cm]{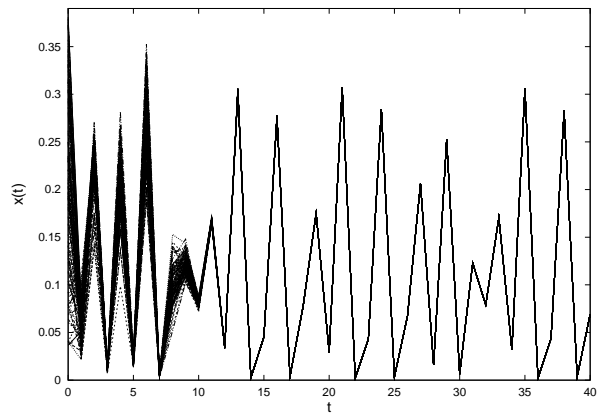}
\end{center}
\caption{\label{Fig.9} Dynamics of a random directed network of 100
sigmoidal neurons with both positive and negative weights, showing
synchronization to a chaotic trajectory starting from random initial
conditions. Parameter values are as in fig.~\ref{Fig.5}.}
\end{figure}

As a second model of a neuronal network we consider a sigmoidal
neuron dynamics with $f(x) = \sigma_\alpha(x)$. In this case the
neuron behaves like one with bias term. We take the interactions
between the neurons to be also given by a sigmoidal function $g(x) =
\sigma_\beta(x)$. The resulting synchronized dynamics satisfies
\begin{equation} \label{3} s(t+1) = \sigma_\alpha(s(t)) +
\epsilon \sigma_\beta(s(t)).\end{equation} For the special case
$\epsilon = -1$, the dynamics  of eq.~(\ref{3}) has been
analytically shown to be chaotic if $\alpha>2\beta$ \cite{Wang91}.
Here we consider a whole range of $\epsilon$-values.
 The bifurcation diagram of eq.~(\ref{3}) is plotted in
fig.~\ref{Fig.3}, for a set of parameter values. It is seen that the
dynamics has a complicated dependence on $\epsilon$, and there are
many regions of chaotic behavior interspersed with periodic windows.
In fig.~\ref{Fig.5} the mixed transverse exponent $\chi$ is plotted
as a function of the complex eigenvalue $\lambda$, where the blue
color shows regions of synchronization.
Figure \ref{Fig.9} shows the onset of synchronization to chaos in a
random directed network of 100 sigmoidal neurons, where the
probability of a directed link from a vertex to another is taken to
be 0.25 for a positive link and 0.01 for a negative link.
As in the leaky neuron model, monotonic individual dynamics is
replaced by collective chaotic behavior, this time in a random
directed network having both excitatory and inhibitory links.
 By adjusting the global coupling strength $\epsilon$,
one can observe a wide variety of synchronized dynamical behavior.


\section{Suppression of chaos via synchronization}
Besides the emergence of chaos in networks of simple units, our
theory can also be used to show the possibility of
 simple synchronous dynamics in a network
of chaotic units. In other words, chaos is replaced by simpler
behavior in the network. The field of chaos control is extensive and
includes several well-established methods; for an overview see
\cite{Schoell07} and the references therein. In our case, the
network achieves chaos suppression through synchronization of its
units.

As an example we study chaos suppression in a network of coupled
chaotic logistic maps. It is well-known that the logistic map
\begin{equation}\ell_\rho(x) = \rho x(1-x), \;\; \rho\in[0,4] \mbox{ and } x\in
[0,1],\end{equation} undergoes a period doubling route to chaos as
the parameter $\rho$ is increased from 0 to 4 \cite{Devaney89}. In
the sequel we will consider two different values for the parameter
$\rho$. For $\rho = 2.5$ the logistic map possesses an attracting
fixed point and the Lyapunov exponent is given by $\mu_{\ell_{2.5}}
=-\ln2$. Thus $\ell_{2.5}(x)$ is dynamically simple. On the other
hand for $\rho = 4$ the logistic map is maximally chaotic with a
Lyapunov exponent $\mu_{\ell_4} = \ln 2$.

Consider a network of chaotic logistic maps, with $f(x)
=g(x)=4x(1-x)$ and $\epsilon = -3/8$. In this case the synchronous
solution is given by $s(t) =\ell_{2.5}(t)$. So the whole
synchronized network displays simple dynamical behavior, although
all units in the network are chaotic. It follows from (\ref{c*}) and
(\ref{r*}) that the network synchronizes if all eigenvalues
$\lambda_k$, for $k\geq 2$, are contained in
$\mathcal{D}(-5/3,10/3)$.

\section{Synchronization condition without eigenvalue calculations}
As we have seen, the synchronous solution can be simple although all
units of the network are chaotic. In this case it is possible to
state a sufficient condition for synchronization without the
explicit calculation of the Laplacian eigenvalues.  It follows from
Gershgorin's disk theorem \cite{Horn90} that all eigenvalues of
$\mathcal{L}$ are contained in $\mathcal{D}(1,r)$,
where\footnote{Clearly $r\geq1$, and equality holds if and only if
the weights are all nonnegative or all nonpositive. Note that $r$
can be much larger than 1 if there exist vertices in the graph with
small in-degree ($d_i \ll 1$), due to cancellations of positive and
negative weights.}
\begin{equation}\label{r}r := \max_i \frac{\sum_j |w_{ij}|}{|\sum_j w_{ij}|}
.\end{equation}
On the other hand, by (\ref{c*}) and (\ref{r*}), the system
synchronizes if all eigenvalues $\lambda_k$, for $k\geq 2$, of
$\mathcal{L}$ are contained in $\mathcal{D}(c^*,r^*)$. Consequently,
a sufficient condition for synchronization is given by
\begin{equation} \label{Disk} \mathcal{D}(1,r)\subset \mathcal{D}(c^*,r^*).\end{equation}
We consider the case where $g(x) = cf(x)$. In \cite{Bauer08b} we
prove that (\ref{Disk}) holds if and only if
\begin{equation}\label{independent}  \mu_{(1+ \epsilon c)f} < \log \left| \frac{1 +
\epsilon c}{1 + |\epsilon c|r}\right|.\end{equation}
Note that (\ref{independent}) can only be satisfied if the resulting
synchronous behavior is not chaotic, since the right-hand-side of
(\ref{independent}) is non-positive. Hence,  if the synchronized
solution $s(t)$ is not chaotic, it is possible to use the spectral
bound $r$, instead of the whole spectrum of $\mathcal{L}$, to give a
sufficient condition for synchronization.
The advantage is that from (\ref{r}) one can immediately estimate
the effect of changing the network weights without lengthy
eigenvalue calculations.

\section{Discussion} In diffusively-coupled
networks, the whole synchronized network displays the same behavior
as any single individual unit; hence, complex behavior cannot emerge
through synchronization of dynamically simple units. In contrast, as
we have shown in this letter, the direct-coupling scheme leads to
new collective dynamical behavior when the network synchronizes. We
have given an analytical condition for synchronization in terms of
the spectrum of the generalized graph Laplacian and the dynamical
properties of the individual units and coupling functions. In
particular, we have shown that synchronous chaotic behavior can
emerge in networks of simple units, and conversely, chaos can be
suppressed in networks of chaotic units through synchronization.
These results represent a further step towards answering a
fundamental question in complexity, namely, how complex collective
behavior emerges in networks of simple units.

The setting presented here allows for studying synchronization in
general network architectures. Such generality is important for
applications because the connection structure of many real-world
networks is unidirectional and the influence of neighboring units
can be excitatory or inhibitory, as in neuronal networks. We have
applied our theoretical findings to two neuronal network models, and
have shown that, by changing a single parameter such as the coupling
constant, the network can exhibit quite a rich range of dynamical
behavior in its synchronized state. The results presented here
provide insight on how new dynamical behavior may be induced in
neuronal networks by changing the synaptic coupling strengths in a
learning process.


\end{document}